\documentclass[]{aa}

\usepackage{txfonts}

\usepackage{graphicx}

\begin{document}

\def\P{\bar{\Phi}}

\def\st{\sigma_{\rm T}}

\def\vk{v_{\rm K}}

\def\sles{\lower2pt\hbox{$\buildrel {\scriptstyle <}
   \over {\scriptstyle\sim}$}}

\def\sgreat{\lower2pt\hbox{$\buildrel {\scriptstyle >}
   \over {\scriptstyle\sim}$}}

\title{Solving the main cosmological puzzles with a 
generalized time varying vacuum energy}

\author{Spyros Basilakos}
\institute{Research Center for Astronomy, Academy of Athens, 
GR-11527 Athens, Greece, 
\email{svasil@academyofathens.gr}}

\titlerunning{Cosmological puzzles and a time varying vacuum}

\date{Received / Accepted }

\abstract
{We study the dynamics of the FLRW flat cosmological models 
in which the vacuum energy density varies with time, $\Lambda(t)$.
In particular, we investigate the dynamical 
properties of a generalized vacuum model and we 
find that under certain circumstances the vacuum term in the
radiation era varies as $\Lambda(z) \propto (1+z)^{4}$, 
while in the matter era 
we have $\Lambda(z) \propto (1+z)^{3}$ up to 
$z\simeq 3$ and $\Lambda(z)\simeq \Lambda$ for $z\le 3$. 
The confirmation of such a behavior would be of paramount importance 
because it could provide a solution  
to the cosmic coincidence problem as well as to the fine tuning
problem, without changing the 
well known (from the concordance $\Lambda$-cosmology) Hubble expansion.

}

\maketitle

\section{Introduction}
The analysis of the available high quality cosmological 
data (supernovae type Ia, CMB, galaxy clustering, etc.)
have converged during the last decade towards a cosmic expansion
history that involves a spatial flat geometry and 
a recent accelerating expansion of the
universe (Spergel et al. 2007; Davis et al. 2007; 
Kowalski et al. 2008; Komatsu et al. 2009 and references therein).
This expansion has been attributed to an energy component
(dark energy) with negative pressure which dominates the universe at
late times and causes the observed accelerating expansion. The
simplest type of dark energy corresponds to the cosmological
constant (see for review Peebles \& Ratra 2003). The so called
concordance $\Lambda$ model fits accurately the current 
observational data and thus it is an excellent candidate to be the 
model which describes the observed universe. 

However, the concordance model suffers from, among others
(cf. Perivolaropoulos 2008),
two fundamental problems: (a)  
{\it the fine tuning problem} ie., the fact that the observed value of the
vacuum density ($\rho_{\Lambda}=\Lambda c^{2}/8\pi G$) 
is more than 120 orders of magnitude below that 
value found using quantum field theory (Weinberg 1989) and (b) 
{\it the coincidence problem} ie., the matter energy density 
and the vacuum energy density are of the same
order prior to the present epoch, despite the fact that the former 
is a function of time but the latter not (Peebles \& Ratra 2003).
Attempts to solve the coincidence problem have been presented in the 
literature (see Egan \& Lineweaver 2008
and references therein), in which
an easy way to overpass the coincidence problem is to replace the
constant vacuum energy with a dark energy that evolves with time.
The simplest approach is to consider a  
tracker scalar field $\phi$ in which it 
rolls down the potential energy $V(\phi)$ and therefore it 
could mimic the dark energy 
(see Ratra \& Peebles 1988; Weinberg 1989; Turner \& White 1997;
Caldwell, Dave \& Steinhardt 1998; Padmanabhan 2003).
Nevertheless, the latter consideration does not really solve the 
problem because the initial value of the dark energy still needs to be
fine tuned (Padmanabhan 2003). Also, despite the fact that the current
observations do not rule out the possibility of a dynamical 
dark energy (Tegmark et al. 2004), they strongly indicate that   
the dark energy equation of state parameter $w\equiv P_{DE}/\rho_{DE}$ 
is close to -1 (Spergel et al. 2007; Davis et al. 2007; 
Kowalski et al. 2008; Komatsu et al. 2009).

Alternatively, more than two decades ago,
Ozer \& Taha (1987) proposed a different pattern in which 
a time varying $\Lambda$ parameter could be a possible 
candidate to solve the two fundamental cosmological puzzles 
(see also Bertolami 1986; Freese et al. 1987; 
Peebles \& Ratra 1988; 
Carvalho, Lima \& Waga 1992; Overduin \& Cooperstock 1998;
Bertolami \& Martins 2000; Opher \& Pellison 2004; 
Bauer 2005; Barrow \& Clifton 2006;
Montenegro \& Carneiro 2007 and references therein). 
In this cosmological paradigm, 
the dark energy equation of state parameter $w$
is strictly equal to -1, but the vacuum energy density (or $\Lambda$) 
is not a constant but  
varies with time. Of course, the weak point in this ideology is the 
unknown functional form of the $\Lambda(t)$ parameter. Also,
in the $\Lambda(t)$ cosmological model there is a coupling 
between the time-dependent vacuum and matter 
Wang \& Meng 2005; Alcaniz \& Lima 2005; 
Carneiro S. et al. 2008; Basilakos 2009; Basilakos, Plionis \& 
Sol\'a 2009).
Indeed, using the combination of the conservation of the total energy
with the variation of the vacuum energy, one can prove that 
the $\Lambda(t)$ model provides either a particle production process
or that the mass of the dark matter particles increases (Basilakos
2009 and references therein).
Despite the fact that 
most of the recent papers in dark energy studies are based 
on the assumption that the dark energy evolves 
independently of the dark matter, 
the unknown nature of both dark matter and dark energy 
implies that at the moment we can not exclude the possibility of 
interactions in the dark sector 
(eg., Zimdahl, Pav\'on \& Chimento 2001; 
Amendola et al. 2003; Cai \& Wang 2005; Binder \& Kremer 2006; Das, 
Corasaniti, \& Khoury 2006; Olivares, 
Atrio-Barandela \& Pav\'on 2008 and references therein).

The aim of this work is along the same lines, attempting 
to generalize the main cosmological properties of the traditional 
$\Lambda$-cosmology by introducing 
a time varying vacuum energy and specifically to 
investigate whether such models can yield a late
accelerated phase of the cosmic expansion, 
without the need of the required, in the classical $\Lambda$-model, 
extreme fine tuning.
The plan of the paper is as follows. 
The basic theoretical elements of the problem are 
presented in section 2, 3 and 4, by solving analytically [for a spatially flat 
Friedmann-Lemaitre-Robertson-Walker (FLRW) geometry]
the basic cosmological equations. Also in these sections we prove that
the concordance $\Lambda$-cosmology is as a particular solution
of the $\Lambda(t)$ models.
In section 5 we place constraints on the main parameters of our model by
performing a likelihood analysis utilizing the recent Union08 SnIa data 
(Kowalski et al. 2008). Also, in section 5 we compare 
the different time varying vacuum 
models with the traditional $\Lambda$
cosmology. In this section we also treat analytically, 
the basic cosmological puzzles (the fine
tuning and the cosmic coincidence problem) with the aid of 
the time varying $\Lambda(t)$ parameter. 
Finally, in section 6 we draw our conclusions.

\section{The time dependent vacuum in the Expanding Universe}
In the context of a spatially flat FLRW geometry the basic 
cosmological equations are:
\begin{equation}
\rho_{tot}=\rho_{f}+\rho_{\Lambda}=3H^{2}
\label{frie1} 
\end{equation}
and 
\begin{equation}
\frac{d({\rho}_{f}+\rho_{\Lambda})}{dt}+3H(\rho_{f}+P_{f}+
\rho_{\Lambda}+P_{\Lambda})=0 \;\;,
\label{frie2} 
\end{equation}
where $\rho_{f}$ is the density of the ''cosmic'' fluid: 
\begin{equation}
\rho_{f}(t)=\left\{ \begin{array}{cc}
       \rho_{m}(t) &
       \mbox{Matter era}\\
       \rho_{r}(t) & \mbox{Radiation era}
       \end{array}
        \right.
\end{equation}
and 
\begin{equation}
P_{f}(t)=\beta\rho_{f}=\left\{ \begin{array}{cc}
       0   &
       \mbox{Matter era} \;\;\beta=0 \\
       \frac{\rho_{r}}{3} & \mbox{Radiation era} \;\;\beta=1/3
       \end{array}
        \right.
\end{equation}
is the corresponding pressure. Also
$\rho_{\Lambda}$ and $P_{\Lambda}$ denote the 
density and the pressure of the vacuum component respectively.
From a cosmological point 
of view, at an early enough epoch, the above generalized cosmic fluid behaves 
like radiation $P_{f}=P_{r}=\rho_{r}/3$ ($\beta=1/3$), then behaves as 
matter $P_{f}=P_{m}=0$ ($\beta=0$) and as long as 
$P_{\Lambda}=-\rho_{\Lambda}$ it creates 
an accelerated phase of the cosmic expansion (see below). 
Notice, that in order to simplify our formalism we use geometrical units 
($8\pi G=c\equiv 1$) in which $\rho_{\Lambda}=\Lambda$.
In the present work, we would like to investigate the potential
of a time varying $\Lambda=\Lambda(t)$ parameter
to account for the observed
acceleration of the expansion of the Universe.
Within this framework it is interesting to mention 
that the equation of state takes the usual form of 
$P_{\Lambda}(t)=-\rho_{\Lambda}(t)=-\Lambda(t)$ [see Ozer \& Taha
  1987; Peebles \& Ratra 1988].
Also, introducing in the global dynamics the idea
of the time-dependent vacuum, 
it is possible to explain the physical properties of the 
dark energy as well as the fine tuning and the coincidence 
problem respectively (see sections 5.1 and 5.2). Using now  
eq.(\ref{frie2}), we have the following useful formula:
\begin{equation}
\dot{\rho_{f}}+3(\beta+1) H\rho_{f}=-\dot{\Lambda}
\label{frie33} 
\end{equation}
and considering eq.(\ref{frie1}) we find:
\begin{equation} 
\dot{H}+\frac{3(\beta+1)}{2} H^{2}=\frac{\Lambda}{2}
\label{frie34} 
\end{equation}
where the over-dot denotes derivatives with respect to time.
If the vacuum term is negligible, $\Lambda(t) \longrightarrow 0$, then 
the solution of the above equation reduces to $H(t)=2(\beta+1)^{-1}/3t$.
Therefore, in the case of $\beta=0$ (matter era) 
we get the Einstein de-Sitter model as we should, $H(t)=2/3t$, while 
for $\beta=1/3$ we trace the radiation phase of the Universe ie., 
$H(t)=1/2t$.
On the other hand, if we consider the case of $\Lambda(t)\ne 0$ then 
it becomes evident (see eq.\ref{frie33}) that there is a coupling between 
the time-dependent vacuum and matter (or radiation) component. 

Of course, in order to solve the above differential equation we need to 
define explicitly the functional form of the $\Lambda(t)$ component. 
Note, that the traditional $\Lambda=const$ cosmology 
can be described directly by the integration of the eq.(\ref{frie34})  
[for more details see section 3.1]. 

It is worth noting that the $\Lambda(t)$ scenario has the caveat of its unknown exact
functional form, which however is also the case for the vast majority 
of the dark energy models.
Therefore, in the literature
there have 
been different phenomenological parametrizations which treat 
the time-dependent $\Lambda(t)$ function.
In particular, Freese et al. (1987) considered that 
$\Lambda(t)=3c_{1}H^{2}$, with the constant $c_{1}$ being the ratio of the 
vacuum to the sum of vacuum and matter density (see
also Arcuri \& Waga 1994). Chen \& Wu (1990) proposed a different
ansatz in which $\Lambda(t) \propto a^{-2}$.

Recently, many authors (see for example 
Ray, Mukhopadhyay \& Meng 2007; Sil \& Som 2008 and references therein)
have investigated the global dynamical properties of the universe
considering that the vacuum energy density decreases linearly 
either with the energy density or with the square Hubble parameter. 
Attempts to provide a theoretical explanation for 
the $\Lambda(t)$ have also been presented in the 
literature (see 
Shapiro \& Sol\'a 2000; Babi\'c et al. 2002; Grande et al. 2006; Sol\'a 2008 
and references therein). These 
authors found that a time dependent vacuum could
arise from the renormalization group (RG) in quantum field theory. 
The corresponding solution for a running vacuum 
is found to be $\Lambda(t)=c_{0}+c_{1}H^{2}(t)$ [where $c_{0}$ and
  $c_{1}$ are constants; Grande et al. 2006] 
and it can mimic the quintessence or phantom 
behavior and a smoothly transition between the two.
Alternatively, Schutzahold (2002) used 
a different pattern in which the vacuum term is proportional to
the Hubble parameter, $\Lambda(a) \propto H(a)$ [see
also Carneiro et al. 2008], while Basilakos (2009) considered
a power series form in $H$. It is worth noting, that the
linear pattern, $\Lambda(a) \propto H(a)$, has been motivated theoretically
through a possible connection of cosmology with
the QCD scale of strong interactions (Schutzhold 2002).
In this context, it has also been proposed that the 
vacuum energy density can be defined from a possible link 
of dark energy with QCD and
the topological structure of the universe (Urban \& Zhitnitsky 2009).

In this paper we have phenomenologically 
identified a functional form of $\Lambda(a)$ 
for which we can solve the main differential equation 
(see eq.\ref{frie34}) analytically. This is:
\begin{equation}
\Lambda_{\gamma m}(t)=3\gamma H^{2}(t)+2mH(t)+3n(\beta+1-\gamma){\rm e}^{2mt} 
\label{frie444} 
\end{equation}
where the constants $m$ and $n$ are included 
for consistency of units (see below).
Although, the above functional form was not
motivated by some physical theory, but rather 
phenomenologically by the fact that it provides analytical solutions
to the Friedmann equation, its exact form can be physically 
justified {\em a posteriori}
within the framework of the previously mentioned theoretical 
models (see appendix A).

Using now eq.\ref{frie444}, the generalized 
Friedmann's equation (see eq.\ref{frie34}) becomes
\begin{equation}
\dot{H}=-\frac{3(\beta+1-\gamma)}{2}H^{2}+mH+\frac{3n(\beta+1-\gamma)}{2}{\rm e}^{2mt}  
\label{frie344} 
\end{equation}
and indeed, it is routine to perform the integration 
of eq.(\ref{frie344}) to obtain (see appendix B):
\begin{equation}
\label{frie555a} 
H(t)=\sqrt{n}{\rm e}^{mt}{\rm coth}\left[\frac{3(\beta+1-\gamma)\sqrt{n}}{2}S(t)\right] 
\end{equation}
where 
\begin{equation}
S(t)=\left\{ \begin{array}{cc}
        ({\rm e}^{mt}-1)/m & \;\;\;\;m\ne 0 \\
       t & \;\;\;\;m=0
       \end{array}
        \right.
\label{SS} 
\end{equation}
while the range of values for which the above integration
is valid is $n \in (0,+\infty)$ [for negative $n$ values see the appendix]. 
Using now the definition of the Hubble parameter $H\equiv {\dot a}/a$, the 
scale factor of the universe $a(t)$, evolves with time as 
\begin{equation}
a(t)=a_{1}
\sinh^{\frac{2}{3(\beta+1-\gamma)}}
\left[\frac{3(\beta+1-\gamma)\sqrt{n}}{2}S(t)\right] \;\;.
\label{frie565} 
\end{equation}
It is worth noting, that the relevant 
units of $m\ne 0$ should correspond to 
$time^{-1}$, which implies that $m\propto H_{0}$.
The parameter $a_{1}$ is the constant of integration given by
\begin{equation}
a_{1}\equiv \left(\frac{\rho_{f0}}{\rho_{\Lambda 0}}\right)^{\frac{1}{3(\beta+1-\gamma)}}
\label{normden} 
\end{equation} 
where $\rho_{f0}$ and $\rho_{\Lambda 0}$ are the corresponding 
densities at the present time [for which $a(t_{0})\equiv 1$].  

In this context, the density of the cosmic fluid evolves with
time (see eq.\ref{frie1}) as:
\begin{equation}
\rho_{f}(t)=3H^{2}(t)-\Lambda_{\gamma m}(t)
\end{equation}
or
\begin{equation}
\rho_{f}(t)=3(1-\gamma) H^{2}(t)-2mH(t)-3n(\beta+1-\gamma){\rm e}^{mt} \;\;.
\label{den22} 
\end{equation} 
In the following sections, we investigate thoroughly 
whether such a generalized vacuum component 
in an expanding Universe allows
for a late accelerated phase of the Universe 
and under which circumstances such
an approach provides a viable solution 
to the fine tuning problem as well as to the 
cosmic coincidence problem.

\section{The matter+vacuum scenario}
In a matter+vacuum expanding universe
($\rho_{f}\equiv \rho_{m}$), we attempt to investigate 
the correspondence of the $\Lambda(t)$ pattern with the traditional 
$\Lambda$-cosmology in order to show
the extent to which they compare. In particular, 
we will prove that the Hubble expansion, provided 
by the current time-dependent vacuum, is
a generalization to that of the traditional $\Lambda$ cosmology. 
Note, that in the present formalism the matter era 
corresponds to $\beta=0$.

\subsection{The standard $\Lambda$-Cosmology}
Let us first investigate the solution for $(\gamma,m)=(0,0)$.
The vacuum term eq.(\ref{frie444}) of the problem becomes constant and 
is given by $\Lambda_{00}(a)=\Lambda=3n$. 
In this framework, the Hubble function (see eq.\ref{frie555a}) is
\begin{equation}
H_{\Lambda}(t)=\sqrt{\frac{\Lambda}{3}}
\coth\left(\frac{3}{2}\sqrt{\frac{\Lambda}{3}}\;t \right) \;\;.
\label{frie556} 
\end{equation}
Now, using the well know parametrization 
\begin{equation}
\label{aln} 
\Lambda=3n=3H^{2}_{0}\Omega_{\Lambda} \;\;\;\; \Omega_{\Lambda}=1-\Omega_{m} 
\end{equation} 
the scale factor of the universe is given by
\begin{eqnarray} 
\label{all} 
a_{\Lambda}(t)=a_{1}
\sinh^{\frac{2}{3}}\left(\frac{3H_{0}\sqrt{\Omega_{\Lambda}}t}{2}\right) 
\end{eqnarray} 
where (see eq.\ref{normden})  
\begin{eqnarray} 
\label{all1} 
a_{1}=\left(\frac{\rho_{m0}}{\rho_{\Lambda 0}}\right)^{1/3}=
\left(\frac{\Omega_{m}}{\Omega_{\Lambda}}\right)^{1/3} \;\;.
\end{eqnarray} 
The cosmic time is related with the scale factor as
\begin{equation}
t_{\Lambda}(a)=\frac{2}{3\sqrt{\Omega_{\Lambda}}H_{0}  }
{\rm sin^{-1}h} \left(\sqrt{ \frac{\Omega_{\Lambda}} {\Omega_{m}}} 
\;a^{3/2} \right) \;\;.
\end{equation}

Combining the above equations we can define the Hubble expansion as a function 
of the scale factor:
\begin{eqnarray} 
\label{hub1} 
H_{\Lambda}(a)=H_{0}[\Omega_{\Lambda}+\Omega_{m}a^{-3}]^{1/2} \;\;\;.
\end{eqnarray} 
In principle, $H_{0}$ and $\Omega_m$ are constrained by the recent WMAP data 
combined with the distance measurements from the type Ia 
supernovae (SNIa) and the Baryonic Acoustic Oscillations (BAOs) 
in the distribution of galaxies. Following the 
recent cosmological results of (Komatsu et al. 2009), we fix the 
current cosmological parameters as $H_{0}=70.5$km/sec/Mpc and 
$\Omega_{m}=1-\Omega_{\Lambda}=0.27$. 
The current age of the universe ($a=1$) is $t_{0\Lambda}\simeq 13.77$Gyr, 
while the inflection point takes place at 
\begin{eqnarray} 
      \label{infle}
t_{I\Lambda}=\frac{2}{3\sqrt{\Omega_{\Lambda}}H_{0}}
{\rm sin^{-1}h} \left(\sqrt{ \frac{1} {2}} \right) \;,\;\;
a_{I\Lambda}=\left[\frac{\Omega_{m}}{2\Omega_{\Lambda}}\right]^{1/3} \;.
\end{eqnarray}
Therefore, we estimate $t_{I\Lambda}\simeq 0.51t_{0\Lambda}$ 
and $a_{I\Lambda}\simeq 0.56$.

Finally, due to the fact that the traditional $\Lambda$ cosmology 
is a particular solution
of the current time varying vacuum models with $(\gamma,m)$ 
strictly equal to $(0,0)$, 
the constant value $n$ is always   
defined by eq.(\ref{aln}). Therefore, throughout the paper 
all relevant cosmological quantities are parametrized according to 
$n=\Omega_{\Lambda}H^{2}_{0}$.

\subsection{''The general'' $\Lambda(t)$ Model}
In this section, we examine a more general class of vacuum models with 
$(\gamma,m)\ne (0,0)$ (hereafter $\Lambda_{\gamma m}$ model).
The Hubble expansion and the corresponding evolution of the 
scale factor are (see eq.\ref{frie555a} and eq.\ref{frie565}) 
\begin{equation}
\label{frie755} 
H(t)=\sqrt{\Omega_{\Lambda}}\;H_{0}\;{\rm e}^{mt}{\rm coth}
\left[\frac{3(1-\gamma)\sqrt{\Omega_{\Lambda}}H_{0}}{2m}({\rm e}^{mt}-1)\right] 
\end{equation}
and
\begin{equation}
a(t)=a_{1}
\sinh^{\frac{2}{3(1-\gamma)}}
\left[\frac{3(1-\gamma)\sqrt{\Omega_{\Lambda}}H_{0}}{2m}({\rm e}^{mt}-1) \right] 
\label{frie756} 
\end{equation} 
or 
\begin{equation}
t(a)=\frac{1}{m}{\rm ln}\left[1+\frac{2m}{3(1-\gamma)\sqrt{\Omega_{\Lambda}}H_{0}}\;
{\rm sin^{-1}h}\left(\frac{a}{a_{1}}\right)^{3(1-\gamma)/2} \right] \;\;. 
\label{frie656t2} 
\end{equation}
Obviously, if $(\gamma,m) \longrightarrow (0,0)$ 
[or ${\rm e}^{mt}-1\approx mt$]  
then the $\Lambda_{\gamma m}$ model
tends to the traditional $\Lambda$ cosmology, which 
implies that the latter be considered as particular 
solution of the general $\Lambda_{\gamma m}$ model.
Thus, this limit 
together with eq.(\ref{normden}) provide that
\begin{equation}
a_{1}=\left(\frac{\Omega_{m}}{\Omega_{\Lambda}}\right)^{\frac{1}{3(1-\gamma)}} \;\;.
\end{equation}
Taking the above expressions into account, the basic cosmological quantities  
as a function of the scale factor become
\begin{equation}
H(a)=H_{0}\left[1+g(a)\right][\Omega_{\Lambda}+\Omega_{m}a^{-3(1-\gamma)}]^{1/2} 
\end{equation}
and 
\begin{equation}
\Lambda_{\gamma m}(a)=3\gamma H^{2}+2mH+3H^{2}_{0}\Omega_{\Lambda}(1-\gamma)[1+g(a)]^{2}
\label{frie776}  
\end{equation}
where
\begin{equation}
g(a)=\frac{2m}{3\sqrt{(1-\gamma)\Omega_{\Lambda}}H_{0}}\;
{\rm sin^{-1}h}\left(\sqrt{\frac{\Omega_{\Lambda}}{\Omega_{m}}}
\;a^{3(1-\gamma)/2}  \right) \;\;.
\end{equation}

It is worth noting that if we take 
$(\gamma,m)=(0,m)$ 
with $m \ne 0$ (hereafter mild vacuum model or $\Lambda_{0m}$), the 
corresponding Hubble flow becomes:
\begin{equation}
H(a)=\left[1+g(a)\right]H_{\Lambda}(a) \;\;. 
\end{equation}
Therefore, 
as long as the function $g(a)$ takes small values [$g(a)\ll 1$],
the $\Lambda_{0m}$ model has exactly the constant vacuum  
feature due to $H(a) \approx H_{\Lambda}(a)$. 
In this context, utilizing eq.(\ref{frie776}) we simply have
\begin{equation}
\label{frie676} 
\Lambda_{0m}(a)=2mH(a)+3H^{2}_{0}\Omega_{\Lambda}[1+g(a)]^{2} \;\;.
\end{equation} 

Finally, the fact that the vacuum term has units of $time^{-2}$  
implies that the vacuum term is proportional to $H^{2}_{0}$ or the 
constant $m$ has to satisfy the following scaling relation:
$m \propto H_{0}$ (see also section 2). Therefore, 
in the far future the condition 
$m \propto H_{0}\ne 0$ represents a super-accelerated 
expansion of the universe because 
$a(t)\propto {\rm exp}({\frac{\sqrt{\Omega_{\Lambda}}H_{0}{\rm e}^{mt}}{m}})$.

\subsection{''The modified'' $\Lambda$ Model}
In this case we consider $(\gamma,m)=(\gamma,0)$ with $\gamma \ne 0$ 
(hereafter $\Lambda_{\gamma 0}$ model). 
From  eq.(\ref{frie555a}) we can 
easily write the corresponding Hubble flow as a function of time
\begin{equation}
\label{frie455} 
H(t)=\sqrt{\Omega_{\Lambda}}\;H_{0}
\;{\rm coth}\left[\frac{3(1-\gamma)\sqrt{\Omega_{\Lambda}}H_{0}}{2}\;t\right] \;\;. 
\end{equation}
Using now eqs.(\ref{SS}, \ref{frie565}), the 
scale factor of the universe $a(t)$, evolves with time as 
\begin{equation}
a(t)=a_{1}
\sinh^{\frac{2}{3(1-\gamma)}}
\left[\frac{3(1-\gamma)\sqrt{\Omega_{\Lambda}}H_{0}}{2}\;t\right] 
\label{frie456} 
\end{equation}
where
\begin{eqnarray} 
\label{all2} 
a_{1}=\left(\frac{\Omega_{m}}{\Omega_{\Lambda}}\right)^{1/3(1-\gamma)} \;\;.
\end{eqnarray} 
Inverting eq.(\ref{frie456}) we estimate the cosmic time:
\begin{equation}
t(a)=\frac{2}{3(1-\gamma)\sqrt{\Omega_{\Lambda}}H_{0}  }
{\rm sin^{-1}h} \left(\sqrt{ \frac{\Omega_{\Lambda}} {\Omega_{m}}} 
\;a^{3(1-\gamma)/2} \right) \;\;.
\label{frie456t} 
\end{equation}
The corresponding inflection point [$\ddot{a}(t_{I})=0$] is found to be 
\begin{equation} 
      \label{inflemod}
t_{I}=\frac{2}{3(1-\gamma)\sqrt{\Omega_{\Lambda}}H_{0}}
{\rm sin^{-1}h} \left(\sqrt{ \frac{1-3\gamma} {2}} \right) 
\end{equation}
or
\begin{equation} 
a_{I}=\left[\frac{(1-3\gamma)\Omega_{m}}{2\Omega_{\Lambda}}\right]^{1/3(1-\gamma)} 
\end{equation}
which implies that the condition 
for which an inflection point is present in the evolution of 
the scale factor is $\gamma<1/3$.

As expected, for $\gamma \ll 1$ the 
above solution tends to the concordance model, 
$a_{\gamma 0}(t) \longrightarrow a_{\Lambda}(t)$.
Now from eqs.(\ref{frie455}, \ref{frie456}), using the well known 
hyperbolic formula ${\rm coth^{2}}x-1=1/{\rm sinh^{2}x}$, we have after some algebra that:
\begin{equation}
H(a)=H_{0}[\Omega_{\Lambda}+\Omega_{m}a^{-3(1-\gamma)}]^{1/2} \;\;.
\end{equation}
From this analysis, it becomes 
clear that the Hubble expansion predicted by the 
$\Lambda_{\gamma 0}$ model extents well that of the usual $\Lambda$ cosmology. 
To this end, utilizing eq.(\ref{frie776}) we can obtain the vacuum 
energy density
\begin{equation}
\label{frie476} 
\Lambda_{\gamma 0}(a)=3\gamma H^{2}(a)+
3\Omega_{\Lambda}H^{2}_{0}(1-\gamma) \;\;.
\end{equation}

As we have previously 
mentioned in section 2, the above phenomenological functional form 
(see eq.\ref{frie476}) 
is motivated theoretically by the renormalization group (RG) in the quantum 
field theory (Shapiro \& Sol\'a 2000; Babi\'c et al. 2002; Sol\'a
2008). Moreover, recent studies (see Grande et al. 2006 and 
Grande, Pelinson \& Sol\'a 2009) find that this solution 
alleviates the cosmic coincidence problem (see section 5.1). 
To conclude, it is worth noting that at late 
enough times ($a\gg 1$) the above solution 
asymptotically reaches the de-Sitter regime $\Lambda\sim H^{2}$, 
as far as the global dynamics is concerned.

\section{The radiation+vacuum scenario}
In this section, we consider a universe that is a spatially flat but contains 
both radiation and a time vacuum term.
This crucial period in the cosmic history corresponds to 
$\beta=1/3$.
Therefore, for clarity reasons in the following sections we re-formulate our approach 
by using $\rho_{f}\equiv \rho_{r}$ and $P_{f}\equiv \rho_{r}/3$. These restrictions imply that 
$$\frac{\rho_{f 0}}{\rho_{\Lambda 0}}\equiv 
\frac{\rho_{r 0}}{\rho_{\Lambda 0}}=\frac{\Omega_{r}}{\Omega_{\Lambda}}$$ 
where, $\Omega_{r}\simeq 10^{-4}$ is the radiation density parameter at the 
present epoch derived by the CMB data (see Komatsu et al. 2009).
Within this context, based on eqs.(\ref{frie444}), (\ref{frie565}) and 
(\ref{normden}) we present briefly the following cosmological situations:

\begin{itemize}
\item {\bf radiation+constant vacuum:} $(\gamma,m)=(0,0)$: 
The scale factor is  
\begin{equation}
a(t)=\left(\frac{\Omega_{r}}{\Omega_{\Lambda}}\right)^{\frac{1}{4}}\sinh^{\frac{1}{2}}
\left(\sqrt{\Omega_{\Lambda}}H_{0}t\right) \;\;.
\end{equation} 
Owing to the fact that in this period $t\ll 1$, the above solution reduces to the 
following simple analytic approximation: 
\begin{equation}
\label{approxrad}  
a(t)\approx (2\sqrt{\Omega_{r}}H_{0}t)^{1/2} \;\;\;{\rm with}\;\;\;
H(t)\equiv \frac{\dot{a}}{a}\approx \frac{1}{2t} \;\;.
\end{equation}

\item {\bf radiation+general vacuum:} $(\gamma,m)\ne (0,0)$: 
This general scenario provides 
\begin{equation}
a(t)=\left(\frac{\Omega_{r}}{\Omega_{\Lambda}}\right)^{\frac{1}{4\gamma_{1}}}
\sinh^{\frac{1}{2\gamma_{1}}}
\left[\frac{2\gamma_{1}\sqrt{\Omega_{\Lambda}}H_{0}}{m}({\rm e}^{mt}-1) \right]
\end{equation}
where $\gamma_{1}=1-3\gamma/4$. 
The vacuum component as a function of time (see eq.\ref{frie444}) is
\begin{equation}
\Lambda_{\gamma m}(t)\approx \frac{4(1-\gamma_{1})}{4\gamma^{2}_{1}t^{2}}+\frac{m}{\gamma_{1}t}
\label{llapprox1}  
\end{equation}
or
\begin{equation}
\label{llapprox2}  
\Lambda_{\gamma m}(a)\approx \frac{4(1-\gamma_{1})\Omega_{r}H^{2}_{0}}{a^{4\gamma_{1}}} 
+\frac{2m\sqrt{\Omega_{r}}H^{2}_{0}}{a^{2\gamma_{1}}} 
\;\;.
\end{equation}
It is very interesting the fact that during the radiation epoch
$\Lambda_{\gamma m}(a) \propto a^{-4\gamma_{1}}$. For small values 
of $\gamma$ or $\gamma_{1}\simeq {\cal O}(1)$, the latter relation 
implies that as long as the scale factor tends to zero the 
vacuum term goes rapidly to infinity 
(see section 6).
In the case of $(\gamma,m)=(0,m)$ [or $\gamma_{1}=1$], the vacuum term 
(see eqs.\ref{llapprox1} and \ref{llapprox2}) 
varies with time as 
\begin{equation}
\Lambda_{0m}(t)\approx \frac{m}{t}\approx \frac{2m\sqrt{\Omega_{r}}H^{2}_{0}}{a^{2}} \;\;.
\end{equation}
Now the vacuum component evolves as $\Lambda_{\gamma 0}(a) \propto
a^{-2}$, in agreement with the Chen \& Wu (1990) model.

\item {\bf radiation+modified vacuum:} $(\gamma,m)=(\gamma,0)$, $\gamma\ne 0$: 
In this cosmological model we have
\begin{equation}
a(t)=\left(\frac{\Omega_{r}}{\Omega_{\Lambda}}\right)^{\frac{1}{4\gamma_{1}}}
\sinh^{\frac{1}{2\gamma_{1}}}
\left[2\gamma_{1}\sqrt{\Omega_{\Lambda}}H_{0}\;t \right]
\end{equation}
where $\gamma_{1}=1-3\gamma/4$. The approximate solution now becomes 
\begin{equation}
\label{approxrad1}  
a(t)\approx (2\gamma_{1}\sqrt{\Omega_{r}}H_{0}t)^{1/2\gamma_{1}} \;\;\;{\rm with}\;\;\;
H(t)\approx \frac{1}{2\gamma_{1}t} \;\;.
\end{equation}
The vacuum component (see eq.\ref{frie444}) evolves with time as
\begin{equation}
\Lambda_{\gamma 0}(t)\approx \frac{4(1-\gamma_{1})}{4\gamma^{2}_{1}t^{2}}
\end{equation}
or
\begin{equation}
\Lambda_{\gamma 0}(a)\approx \frac{4(1-\gamma_{1})\Omega_{r}H^{2}_{0}}{a^{4\gamma_{1}}}
\simeq \Lambda_{\gamma m}(a)\;\;. \;\;.
\end{equation} 
Obviously, for $a\longrightarrow 0$ 
[$\gamma_{1}\simeq {\cal O}(1)$]
the vacuum energy density 
goes rapidly to infinity.

\end{itemize}

\begin{figure}
\includegraphics[angle=0,scale=0.45]{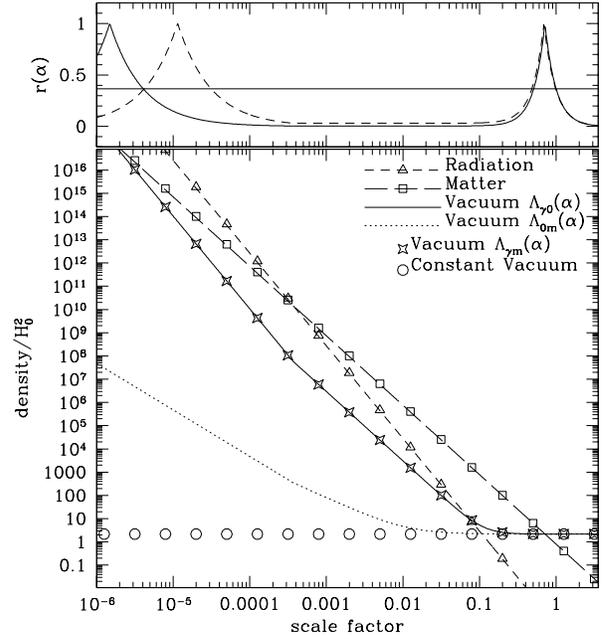}
\caption{{\it Upper Panel:} The evolution of the proximity parameter 
for the $\Lambda_{\gamma 0}$ cosmological model. 
Note, that the scale factor is normalized to unity at the present time.
The lines correspond to $\gamma=0.004$ (solid) and $\gamma=0.03$
(dashed).
{\it Bottom Panel:} The evolution of the radiation, matter and vacuum density
considering different kind of vacuums
(after fitting the constants using the 
Union08 SnIa data and $\Omega_{m}=0.27$,
$H_{0}=70.5$Km/s/Mpc). 
I) traditional $\Lambda$-cosmology: 
radiation density (open triangles), matter density (open
squares) and constant vacuum density (open circles). 
II) modified $\Lambda$-cosmology, $\gamma \ne 0$, $\Lambda_{\gamma 0}$:  
radiation density (dashed line), matter density (long-dashed line)
and vacuum density (solid line). III) The evolution of the 
mild vacuum, $m\ne 0$, $\Lambda_{0m}$ and IV) the evolution of the general 
vacuum, $\Lambda_{\gamma m}$ (open stars).} 
\label{fig1}
\end{figure}

\begin{table}
\caption[]{Numerical results. The $1^{st}$ column 
indicates the vacuum model used 
(the last two rows correspond to the fine tuning problem). 
Note, that the basic cosmological parameters were taken to be 
$\Omega_{m}=0.27$ and $H_{0}=70.5$Km/sec/Mpc.
Finally, the current age of the universe $t_{0}$ has units of Gyr.}
\tabcolsep 9pt
\begin{tabular}{cccccc} 
\hline
Model & $\gamma$ & $m/H_{0}$ & $t_{0}$&$\frac{\Lambda(t_{inf})}{\Lambda(t_{0})}$ 
&$\frac{\Lambda(t_{pl})}{\Lambda(t_{0})}$\\ \hline \hline 
$\Lambda$ & 0& 0 & 13.77&1&1\\
$\Lambda_{\gamma 0}$ & $0.004$& 0 &
13.82&$10^{102}$&$10^{124}$\\
$\Lambda_{0m}$ & 0& $2.4\times 10^{-3}$ &
13.75&$10^{51}$&$10^{63}$\\
$\Lambda_{\gamma m}$ & $0.004$&$2.8\times 10^{-3}$ &
13.80&$10^{102}$&$10^{124}$\\
\end{tabular}
\end{table}

\section{Tackling the Cosmological puzzles}
As we have stated already in the introduction, there is a possibility 
for the vacuum energy to be a function of time 
rather than having a constant value. Therefore, in this section
we compare the cosmic phases of the 
$\Lambda(t)$ scenarios (described in the previous sections) 
and the concordance $\Lambda$-cosmology. 
The aim here is to investigate the consequences
of such a comparison on the basic cosmological puzzles
namely cosmic coincidence problem and fine tuning problem.

\subsection{The coincidence problem}
In order to investigate the coincidence problem we 
define the time-dependent proximity parameter of 
$\rho_{m}(a)$ (see eq.\ref{den22}) and 
$\rho_{\Lambda}(a)$ [see Egan \& Lineweaver 2008 and references therein]:
\begin{equation}
\label{prox} 
r(a)\equiv {\rm min}\left[\frac{\rho_{\Lambda}(a)}{\rho_{m}(a)},
\frac{\rho_{m}(a)}{\rho_{\Lambda}(a)} \right] \;\;.
\end{equation}
Note, that in this work we use $\rho_{\Lambda}(a)\equiv \Lambda(a)$ 
[see eq.\ref{frie444}].
If the two densities differ by many orders of magnitude then 
$r\simeq 0$. On the other hand if the two densities are equal the proximity
parameter is $r=1$. The current observational data shows 
that the proximity parameter
at the present time ($a=1$) is
$r_{0}=\frac{\rho_{m}(1)}{\rho_{\Lambda}(1)}
=\frac{\Omega_{m}}{\Omega_{\Lambda}}\simeq 0.37$. Therefore, a 
cosmological model may suffer from the so 
called coincidence problem 
if its proximity parameter is close to zero before the inflection 
point, $r(a<a_{I})\sim 0$.
As an example, for the traditional $\Lambda$-cosmology we
have $r(a<0.56)\sim 0$. In contrast, if for a particular model
we find that $r(a<a_{I})={\cal O}(1)$ then 
this model possibly does not suffer from 
the cosmic coincidence problem.

In particular, 
suppose that we have a cosmological model which 
accommodates a late time accelerated expansion and it
contains $n$-free parameters, described by the vector 
${\vec \epsilon}=(\epsilon_{1},\epsilon_{2},...,\epsilon_{n})$.
The main question that we should address here is 'what is the 
range of input $(\epsilon_{1},\epsilon_{2},...,\epsilon_{n})$
parameters for which the coincidence problem 
can be avoided?' Below we implement the following tests.

(i) We find the range of the free parameters of the considered
cosmological model that implies $r\simeq r_{0}$ for at least two different 
epochs, one of which is precisely the present epoch. 

(ii) We know that for epochs between the inflection point and the
present time $a_{I}\le a \le 1$, the proximity parameter is
$r(a)\ge r_{0}$. As an example, for the traditional $\Lambda$-cosmology we
have $r(a)\ge 0.37$. Thus, the goal here is to define the range of the
free parameters in which at least a second region 
with $r(a<a_{I})\ge r_{0}$ occurs before the inflection point ($a<a_{I}$).

(ii) Once, steps (i) and (ii) are accomplished, we finally check 
whether the remaining parameters fit the recent SnIa data, by 
performing a standard $\chi^{2}$ minimization. In this work, 
we use the so called Union08 sample of 307 supernovae of 
Kowalski et al. (2008).
In particular, the $\chi^{2}$ function can be written as:
\begin{equation}
\label{chi22} 
\chi^{2}({\vec \epsilon})=\sum_{j=1}^{307} \left[ \frac{ {\cal \mu}^{\rm th}
(a_{j},{\vec \epsilon})-{\cal \mu}^{\rm obs}(a_{j}) }
{\sigma_{j}} \right]^{2} \;\;.
\end{equation}
where $a_{j}=(1+z_{j})^{-1}$ is the observed scale factor of
the universe, $z_{j}$ is the observed redshift, ${\cal \mu}$ is the 
distance modulus ${\cal \mu}=m-M=5{\rm log}d_{\rm L}+25$
and $d_{\rm L}(a,{\vec \epsilon})$ is the luminosity distance, given by 
\begin{equation}
d_{\rm L}(a,{\vec \epsilon})=\frac{c}{H_{0}a} \int_{a}^{1} \frac{dx}{x^{2}E(x)} \;\;,
\end{equation}
where ${\vec \epsilon}$ is the vector containing the unknown free
parameters and
$c$ is the speed of light ($\equiv 1$ here). 

A cosmological model for which the present tests are successfully passed
should not suffer of the coincidence problem.
Below, we apply our tests 
to the current $\Lambda(t)$ cosmological models (see also Table 1).
\begin{itemize}
\item The modified vacuum model with ${\vec \epsilon}=(\gamma,0,...0)$: 
We sample the unknown $\gamma$ parameter as follows: $\gamma \in
(-1,1/3)$ in steps of $10^{-4}$. 
We confirm that in the range of $\gamma \in [0.004,0.03]$ 
the $\Lambda_{\gamma 0}$ model\footnote{Note, that from a theoretical
  viewpoint the predicted value of the $\gamma$ parameter is 
$|\gamma|=\frac{1}{12\pi}\,\frac{M^2}{M_P^2}$, 
where $M_P$ is the Planck mass and $M$ is an effective mass
parameter representing the average mass of the heavy particles of
the Grand Unified Theory (GUT) near the Planck scale, after taking
into account their multiplicities. In the case of $M\sim M_{P}$
we can derive an upper limit of $|\gamma| \le 1/12\pi$
(for more details see Basilakos et al. 2009).} 
satisfies both the criterion (i) and
(ii) respectively. Also, we verify that this range of values fits 
very well the SnIa data,
$\chi^{2}_{min}/{\rm dof}\simeq 1.01$. Notice, that for
$\gamma>0.03$ the criterion (i) is not satisfied.
As an example, in the upper panel of figure 1 we present 
the evolution of the proximity parameter 
for $\gamma=0.004$ (solid line) and $0.03$ (dashed line).
It is worth noting, that 
for $0.1\le a \le 0.34$ (or $2\le z \le 10$) the vacuum density 
is low enough ($r\sim 0$) to allow galaxies and galaxy clusters to
form (Garriga, Livio \& Vilenkin 1999; Basilakos et al. 2009).
From now on, we will utilize $\gamma\simeq 0.004$ that corresponds to the best 
fit parameter. To this end it becomes clear that 
the $\Lambda_{\gamma 0}$ model passes the above criteria and
thus it does not suffer form the cosmic coincidence problem.

\item The mild vacuum model with ${\vec \epsilon}=(0,m,...0)$:  
In this cosmological model, we find that for $m \ge 0.17H_{0}$, 
the corresponding age of the universe is $t_{0}\le 12.7$Gyr. The latter 
appears to be ruled out by the ages of the oldest 
known globular clusters (Krauss 2003; Hansen et al. 2004).
Using this constrain the unknown $m$ parameter 
has an upper limit of $0.17H_{0}$ and thus we perform the
following sampling: 
$m \in [5\times 10^{-4}H_{0},0.17H_{0})$ in steps of $5\times
10^{-4}H_{0}$. Within this range,  
we find that the required (i) and (ii) criteria are not satisfied. 
Thus, the $\Lambda_{0m}$ cosmological model suffers of the 
coincidence problem. The resulting minimization provides:
$m=2.4^{+6}_{-1}\times 10^{-3}H_{0}$ with 
$\chi^{2}_{min}/{\rm dof}\simeq 1.01$.
Note that the errors of the fitted 
parameters represent $1\sigma$ uncertainties.

\item The general vacuum model with ${\vec \epsilon}=(\gamma,m,...0)$: 
This vacuum cosmological model contains 2 free parameters. Using the previous mentioned 
sampling, we obtain that our main criteria for the $\Lambda_{\gamma
  m}$ scenario are full-filled for $\gamma \in [0.004,0.02]$, 
$m \in [1.4\times 10^{-3}H_{0},9\times 10^{-3}H_{0}]$ with
$\chi^{2}_{min}/{\rm dof}\in [1.01,1.02]$.
Throughout the rest of the paper we will use the best fit parameters. These are:
$m\simeq 2.8 \times 10^{-3}H_{0}$ and $\gamma\simeq 0.004$ 
\end{itemize}

In addition to the SnIa data, we further check our statistical 
results using the dimensionless distance to the surface of the last scattering
$R=1.71 \pm 0.019$ (Komatsu et al. 2009), and the baryon acoustic oscillation
(BAO) distance at $z=0.35$, $A=0.469\pm 0.017$ 
(Eisenstein et al. 2005; Padmanabhan, et al. 2007).
We find that the above results remain unaltered.

\subsection{The cosmic evolution - fine tuning problem}
Using now our best fit parameters for the different kind of vacuums, 
we present in figure 1
the corresponding normalized energy densities, vacuum 
$\Lambda(a)/H^{2}_{0}$, matter $\rho_{m}(a)/H^{2}_{0}$ and radiation 
$\rho_{r}(a)/H^{2}_{0}$ as a function of the scale factor.
We verify that both the $\Lambda_{\gamma 0}$ (solid line) and  
$\Lambda_{\gamma m}$ (open stars) solutions are models that provide large 
values for the vacuum energy density at early epochs, in contrast 
with the usual $\Lambda$ cosmology (open circles), 
in which the vacuum energy density remains constant everywhere.
Also, within a Hubble time ($0<a\le 1$) and for each $(\gamma,
m)$ pair, we find the well known cosmic behavior 
for the matter density, $\rho_{m}(a)\propto a^{-3}$ and the radiation 
density, $\rho_{r}(a)\propto a^{-4}$ respectively. 
As an example, in figure 1 we present the density evolution of the cosmic
fluid for the $\Lambda_{\gamma 0}$ cosmological model: matter (dashed line) 
and radiation (dot-dashed line). For comparison we also plot the
predictions of the traditional $\Lambda$ cosmology: 
matter (open squares) and radiation (open triangles). From figure 1,
it becomes clear that the radiation-matter 
equality takes place close to
$a_{rm}\simeq 3.7\times 10^{-4}\simeq \Omega_{r}/\Omega_{m}$. 
For those vacuum models where $m\ne 0$ ($\Lambda_{0m}$ and
$\Lambda_{\gamma m}$), we verify that the 
behavior of their cosmic fluid (matter+radiation) deviates
from the $\Lambda$ solution in the far future ($t\gg t_{0}$), since 
the exponential term ${\rm e}^{mt}$ in eq.(\ref{den22}) plays an important
role in the global dynamics (see section 3.4 and below).

In particular, for the $\Lambda_{\gamma 0}$
vacuum scenario (the same behavior holds for $\Lambda_{\gamma m}$)
we have revealed the following phases:
(a) at early enough times ($\alpha<a_{rm}$) the scale factor of the 
universe tends to its minimum value,   
$a\longrightarrow 0$, which means that the vacuum energy density 
initially goes quickly to infinity. 
So, as long as the scale factor increases 
the vacuum energy rolls down rapidly as $\Lambda_{\gamma 0}(a)\propto
a^{-4\gamma_{1}}$ [where $\gamma_{1}\sim {\cal O}(1)$]. 
This evolution may treat the fine tuning problem. Indeed, 
for $\gamma \in (0,1/3)$, we find that prior to the inflation point 
($t_{inf}\sim 10^{-32}$sec), the vacuum energy density
divided by its present value is 
$\Lambda(t_{inf})/\Lambda(t_{0})\sim 10^{102}$
Finally, if we consider that the functional form of 
$\Lambda(a)\propto a^{-4\gamma_{1}}$ is still valid during the 
Planck time ($t_{pl}\sim 10^{-43}$sec), then 
$\Lambda(t_{pl})/\Lambda(t_{0})\sim 10^{124}$
(see the last rows in Table 1), and (b) in the matter era the vacuum density 
continues to roll down but with a 
different power law $\Lambda_{\gamma 0}(a)\propto
a^{-3(1-\gamma)}$ and it tends to a constant value 
close to $a\sim 0.25$ ($z\sim 3$). Finally,
for $a\ge 0.25$ the vacuum energy density is effectively frozen 
to its nominal value, 
$\Lambda_{\gamma 0}(a)\simeq \Lambda=3\Omega_{\Lambda}H^{2}_{0}$,
which implies that the considered time varying vacuum model explains 
why the matter energy density and the dark energy density are of the same
order prior to the present epoch. 
The moment of radiation-vacuum equality occurs at $a_{rv}\simeq 0.1\simeq
(\Omega_{r}/\Omega_{\Lambda})^{1/4}$. 
Similarly, the moment of matter-vacuum equality takes place at 
$a_{mv}\simeq 0.72\simeq (\Omega_{m}/\Omega_{\Lambda})^{1/3}$. 
From the observational viewpoint,
in order to investigate whether the vacuum 
energy density follows the above evolution, we need a robust 
cosmological probe at redshifts $z\ge 3$. 
In a recent paper (Basilakos et al. 2009), we have investigated 
how realistic it would be to detect differences among the vacuum models.
In particular, we have found that the Sunayev-Zeldovich cluster 
number-counts (as expected from the survey of the South Pole 
Telescope, Staniszewski et al. 2009, and the Atacama Cosmology 
Telescope, Hincks et al. 2009)
indicate that we maybe able to detect significant 
differences among the vacuum models in the redshift range $2.5 \le z \le 3$ 
at a level of $\sim 6-12\%$, which translates in number count
differences, over the whole sky, of $\sim 100$ clusters
(see figure 6 in Basilakos et al. 2009).

\begin{figure}
\includegraphics[angle=0,scale=0.4]{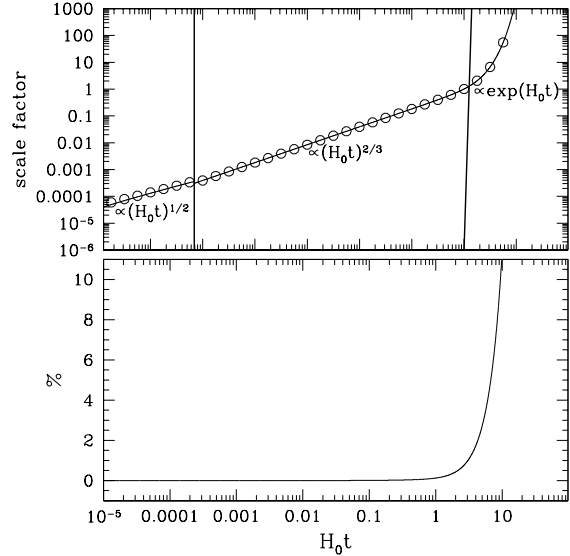}
\caption{{\it Upper Panel:} Comparison of the 
scale factor provided by our $\Lambda_{\gamma 0}$ model 
with the traditional $\Lambda$ cosmology (open points). Note, that we use
$\Omega_{m}=0.27$ and $H_{0}=70.5$Km/s/Mpc model. 
In the bottom panel we present the 
deviation of the scale factors between the 
$\Lambda_{\gamma 0}$ and $\Lambda_{\gamma m}$ model respectively. 
Note, that the scale factor is normalized to unity at the present time.}
\end{figure}

Finally, in figure 1 we also show the evolution of the  
mild vacuum model $\Lambda_{0m}(a)$ (dot line), in which $\gamma=0$. 
Briefly, we get the following dependence:
(a) $\Lambda_{0m}\propto
a^{-2\gamma_{1}}$ for $a<a_{rm}$,
while we estimate that $\Lambda_{0m}(t_{inf})/\Lambda_{0m}(t_{0})
\sim 10^{51}$ and $\Lambda_{0m}(t_{pl})/\Lambda_{0m}(t_{0})
\sim 10^{63}$, (b) between $a_{rm}\le a \le
0.08$ we have $\Lambda_{0m}\propto
a^{-3/2}$ and (c) 
for $a\ge 0.08$ the $\Lambda_{0m}$ becomes constant. 

We would like to end this section with a 
discussion on the evolution of the scale factor.
In particular, our approach provides an evolution of the 
scale factor in the $\Lambda_{\gamma 0}$
model seen in the upper panel of figure 2 as the solid line,
which mimics the corresponding scale factor of the 
$\Lambda$ cosmological model (open points), despite the 
fact that they describe differently the vacuum term.
On the other hand, in the bottom panel of 
figure 2 we present the corresponding deviation
$[(a_{\gamma m}-a_{\gamma 0})/a_{\gamma 0}]\%$, of the growth factors.
It becomes evident, that within the range $0 < H_{0}t< 5$ 
the evolution of the 
scale factor provided by the $\Lambda_{\gamma m}$ model 
closely resembles, the corresponding scale factor of the 
$\Lambda_{\gamma 0}$ model 
(the same result holds also for the $\Lambda$ cosmology).
However, for models where $m\ne 0$, the situation is somewhat different in the far future. 
Indeed, for $H_{0}t\ge 5$ the $\Lambda_{\gamma m}$ (or $\Lambda_{0m}$)
cosmological scenario 
deviates from the $\Lambda_{\gamma 0}$ (or $\Lambda$) model by
$\sim 5-10\%$. Thus, we conclude that the 
models with $m\ne 0$ give a super-accelerated expansion of the
universe in the far future with respect to those vacuum models where $m=0$.

\section{Conclusions}
The reason for which a cosmological constant 
leads to a late cosmic acceleration is because it introduces in 
Friedmann's equation a component which has an equation of state with negative
pressure, $P_{\Lambda}=-\rho_{\Lambda}$. In the last decade the so
called concordance $\Lambda$-cosmology 
is considered to be the model which describes the cosmological
properties of the observed universe because it fits 
accurately the current observational data.
However, the traditional $\Lambda$ cosmology suffers
from two fundamental puzzles. These are the fine tuning 
and the cosmic coincidence problems. An
avenue through which the above cosmological problems 
could be solved is via the time varying vacuum energy
which has the same equation of state as the traditional $\Lambda$-cosmology.

We wish to spell out clearly which are the basic 
assumptions and conclusions of our analysis.
\begin{itemize}
\item We are assuming a time varying vacuum pattern in which 
the specific functional form is: 
$\Lambda(t)=3\gamma H^{2}(t)+2mH(t)+3n(\beta+1-\gamma){\rm e}^{2mt}$,
where $\beta=0$ (matter era) or $\beta=1/3$ (radiation era), 
$n=3\Omega_{\Lambda}H^{2}_{0}$, while the pair $(\gamma,m)$
characterizes the different types of vacuum. It is worth noting, 
that the above functional form includes the effect of the quantum
field theory (for $m=0$) [Shapiro \& Sol\'a 2000; Babi\'c et al. 2002; 
Grande et al. 2006; Sol\'a 2008] and it also extents recent studies
(see for example Ray et al. 2007; Carneiro et al. 2008; Sil \& Som 2008; Basilakos 2009).
Notice, that we can easily prove that the cosmological constant is 
a particular solution of the general vacuum, that $(\gamma,m)=(0,0)$. 
Also we have investigated the following models: (a) modified vacuum in
which $(\gamma,m)=(\gamma,0)$, mild vacuum with $(\gamma,m)=(0,m)$ and 
general vacuum in which $(\gamma,m)\ne (0,0)$.
In this framework, we find that the time evolution of the basic cosmological 
functions (scale factor and Hubble flow) are described in terms of 
hyperbolic functions which can accommodate a late time accelerated
expansion, equivalent to the standard $\Lambda$ model. 

\item We find that that within the framework of either the
modified or general vacuum models the corresponding vacuum term in the
radiation era varies as $\Lambda(a) \propto a^{-4}$ while in the matter
dominated era  
we have $\Lambda(a) \sim a^{-3}$ up to $z=a^{-1}-1 \simeq 3$ 
while $\Lambda(a)\simeq \Lambda=3\Omega_{\Lambda}H^{2}_{0}$ for $z\le 3$.
This vacuum mechanism simultaneously 
sets (a) the value of $\Lambda$ at the present time to its  
observed value and (b) at the Planck time to a value which is 
$10^{124}$ its present value
[$\Lambda(t_{pl})/\Lambda(t_{0})\sim 10^{124}$]. 
Additionally, we verify that our models appear to overcome 
the cosmic coincidence problem. 
Finally, in order to 
confirm the above results, we need to define a robust 
cosmological probe at high redshifts ($z\ge 3$). 
In Basilakos
et al. (2009) we propose that the future 
cluster surveys based on the Sunayev-Zeldovich
detection method give some hope to distinguish the closely resembling
vacuum models at high redshifts.

\end{itemize}

\section*{Appendix}
In this appendix we provide a physical justification  
of the functional form of $\Lambda(a)$ used in our paper.
As we have already mentioned in the section 2, the vacuum 
energy density can take several forms, depending on the 
theoretical approach. Briefly, 
the renormalization group from the quantum field
theory introduces only even powers of $H$ out of which the 
$H^{2}$ is the leading term (Grande et al. 2006; Sol\'a 2008 and
references therein). 
In another vein, the aforementioned possibility that the vacuum energy
could be evolving linearly with $H$ has been motivated theoretically
through a possible connection of cosmology with
the QCD scale of strong interactions (see Schutzhold 2002; 
Carneiro et al. 2008). In this framework, it 
has also been proposed a possible link of dark energy with QCD and
the topological structure of the universe
(Urban \& Zhitnitsky 2009).
The simplest approach therefore to introduce the effects of the 
DE is to consider a potential $V(\phi) \simeq V_{0}+m^{2}\phi^{2}/2$,
where the homogeneous scalar field $\phi$ 
obeys the Klein-Gordon equation. It is well known that 
for $H \simeq const$ the corresponding $\phi$ evolves 
with time as $\phi(t)\simeq \phi_{0}{\rm e}^{mt}$  
(where in general $m$ is a complex number). 
In this context, one would expect that 
the functional form of the $\Lambda(t)$ should contains 
also an additional term of $\phi^{2}(t) \propto {\rm e}^{2mt}$ in order 
to take into account the possible link between dark energy and QCD.

All the above options have merits and demerits. 
In the current paper, the functional form of $\Lambda(t)$ 
is motivated by a combination 
of the above possibilities namely 
$H^{2}(t)$ [RG], $H(t)$ [QCD] and ${\rm e}^{2mt}$ (dark energy).
In particular, the linear combination reads as follows:
$$
\Lambda(t)=n_{1}H^{2}(t)+n_{2}H(t)+n_{3}{\rm e}^{2mt}
$$
which obviously is very similar to the original (phenomenologically selected)
form of $\Lambda(t)$ (equation 7). Finally, from a mathematical
point of view we can select the constants $n_{1}$, $n_{2}$ and $n_{3}$
to match with those presented in the original equation 7.

\section*{Appendix}
With the aid of the differential equation theory we present
solutions that are relevant to our eq.(\ref{frie344}). 
If one is able to have a Riccati differential equation which is given by
the following special form  
\begin{equation}
\frac{dy}{dx}=f(x)y^{2}(x)+my(x)-n{\rm e}^{2mx}f(x)
\label{frie355} 
\end{equation}
then the general solution of eq.(\ref{frie355}) for $n>0$ is
\begin{equation}
y(x)=\sqrt{n}{\rm e}^{mx}{\rm coth}\left[-\sqrt{n}\int_{x_{0}}^{x} 
{\rm e}^{mu}f(u)du \right] \;\;.
\end{equation}
On the other hand, if $n<0$ then the solution of 
eq.(\ref{frie355}) is
\begin{equation}
y(x)=\sqrt{|n|}{\rm e}^{mx}{\rm cot}\left[-\sqrt{|n|}\int_{x_{0}}^{x} 
{\rm e}^{mu}f(u)du \right] \;\;.
\end{equation}
Note, that in our formulation the function $f(x)$ is a constant:
$f(x)=-3(\beta+1-\gamma)/2$. Also, $n<0$ implies that $\Omega_{m}>1$
(or $\Lambda<0$).

\acknowledgements{}
I would like to thank the anonymous referee
for his/her useful comments and suggestions.

\end{document}